# EZFF: Python Library for Multi-Objective Parameterization and Uncertainty Quantification of Interatomic Forcefields for Molecular Dynamics

Aravind Krishnamoorthy[1], Ankit Mishra[1], Deepak Kamal[2], Sungwook Hong[1,3], Ken-ichi Nomura[1,*], Subodh Tiwari[1], Aiichiro Nakano[1], Rajiv Kalia[1], Rampi Ramprasad[2], Priya Vashishta[1]

[1]*Collaboratory of Advanced Computing and Simulations, University of Southern California, Los Angeles, CA 90089-0242*

[2]*Georgia Institute of Technology, 771 Ferst Drive, Northwest Atlanta, Atlanta, GA 30332*

[3]*Department of Physics & Engineering, California State University, Bakersfield, Bakersfield, CA 93311*

Corresponding Email: knomura@usc.edu

**Abstract.**

*Parameterization of interatomic forcefields is a necessary first step in performing molecular dynamics simulations. This is a non-trivial global optimization problem involving quantification of multiple empirical variables against one or more properties. We present EZFF, a lightweight Python library for parameterization of several types of interatomic forcefields implemented in several molecular dynamics engines against multiple objectives using genetic-algorithm-based global optimization methods. The EZFF scheme provides unique functionality such as the parameterization of hybrid forcefields composed of multiple forcefield interactions as well as built-in quantification of uncertainty in forcefield parameters and can be easily extended to other forcefield functional forms as well as MD engines.*

**Keywords:**
*Molecular Dynamics; Interatomic Forcefield; Genetic Algorithm; Global Optimization*

## 1. INTRODUCTION

Molecular Dynamics (MD) is an important technique in computational chemistry, biology and materials science for simulating the structure, dynamics and thermodynamic properties at the atomic scale. While parameter-free *ab initio* quantum molecular dynamics simulations have been successful in simulating atomic dynamics in small (<1000s of atoms) systems over brief timescales (< 100s of ps), modelling realistically complex systems (over $10^6$ atoms) over chemically and biologically relevant timescales (~ μs to ms) requires the use of classical molecular dynamics simulations, where interatomic interactions are approximated by empirical/semi-empirical forcefields, which are sets of parameterized mathematical functions.

The reliability of results from classical MD simulations, and their predictive power, fundamentally depend upon the quality of the forcefields used. Therefore, parameterization of forcefields is a necessary first step in performing high-quality MD simulations. This parameterization process involves the identification of an optimal set of numerical parameters that best approximates experimental or quantum chemical reference data [1, 2] for material systems under investigation [3]. Further, forcefields must be parameterized to simultaneously reproduce several materials properties, necessitating multi-objective optimization techniques. The large number of optimizable empirical parameters (up to several hundred parameters for complex force fields like ReaxFF [4] and COMB [5]) as well as a non-trivial correlation between these variables makes forcefield parameterization a highly complex global optimization problem [6].

Owing to the complexity of handling high-dimensional parameter and objective space, most existing parameterization schemes transform this into more computationally tractable analogues. One of the earliest schemes, the sequential one-parameter parabolic interpolation (SOPPI) [7], casts this as a sequence of one-dimensional local parabolic minimizations, where a single parameter is optimized to minimize a single weighted sum of several objectives. While computationally simple, the SOPPI method has several significant shortcomings, primarily the propensity of the algorithm to converge to a neighboring local minimum, rather than a global minimum as well as poor convergence characteristics if the optimization is started from a poor initial guess [6, 8]. Further, SOPPI is an inherently sequential method that cannot take advantage of vast capabilities of today's highly parallel supercomputers.

These shortcomings are partially addressed in recent multi-objective schemes like GARFfield [9], which use evolutionary algorithms to perform global minimization of a weighted sum of multiple objectives, using an *a priori* user-provided weighting scheme. Other schemes such as Multi-objective evolutionary strategies [10] and MOGA [11] Rotation-invariant Particle Swarm Optimization with isotropic Gaussian Mutation (RIPSOGM) [6] have been developed that evolve the entire Pareto Frontier of multiple forcefield populations at once, without the need to specify *a priori* weights for the different objectives. Existing software frameworks for forcefield optimization are also commonly limited to the parameterization of a single predefined functional forcefield form, such as the Forcefield Toolkit (ffTK) [12] and general automated atomic model parameterization (GAAMP) [13] frameworks for the CHARMM forcefield, Paramfit [14] for AMBER forcefields and MOGA [11] for ReaxFF forcefields. However, to the authors' knowledge, there is no existing general multi-objective global optimization framework that is applicable to parameterization of different forcefield functional forms implemented in different MD engines.

Here, we introduce EZFF, a flexible Python-based multi-objective forcefield optimizer framework for parameterization of multiple forcefield functional forms, including reactive forcefields such as ReaxFF and COMB, using different molecular dynamics engines (LAMMPS [15], GULP [16], RXMD [17] etc.) against multiple user-definable objectives using an *a posteriori* Pareto-dominant multi-objective methods that are proven to be effective for forcefield parameterization [18, 19]. In the next section, we describe the EZFF framework and typical workflow for forcefield parameterization as well as the different objective functions currently supported for force field development. In section III, we illustrate the application of EZFF to the development of several forcefields, including ReaxFF, Stillinger Weber etc.

**2. SOFTWARE DESCRIPTION**

The EZFF source code is written entirely in Python 3 to take advantage of the large user base, and close integration with large number of scientific libraries for data processing, analysis and optimization. Specifically, EZFF makes use of the open source Platypus library [20] for performing evolutionary optimization. Through Platypus, EZFF supports an ensemble of genetic algorithms, including NSGA2 [21], NSGA3 [22], Evolutionary strategies like IBEA, Differential evolution (GDE3), and particle swarm methods like OMOPSO [23] capable of exploring different regions in the parameter space for nonconvex, discontinuous, and multimodal solutions [9], as evident in the several previous studies that have used hard-coded genetic algorithms for forcefield training [24-26].

Figure 1 describes the typical workflow for multi-objective optimization of a classical forcefield using EZFF. The forcefield optimization process begins with three inputs from the user:
1. The functional form of the forcefield to be optimized, as well as the numerical parameters to be determined are provided in a forcefield template file. This forcefield template is identical to a valid forcefield file, where the optimizable parameters are replaced by named variables enclosed in double angular brackets '<< >>'.
2. The user is also required to provide, in a separate file, the maximum and minimum permissible values of these parameters. During global optimization, EZFF generates new forcefields by randomly sampling each variable within the provided minimum and maximum bounds.
3. Finally, the user must provide a set of one or more structural, chemical and energetic properties that the forcefield must reproduce. Deviation from these ground truths values define the objectives (i.e. errors) that must be minimized during the global optimization process.

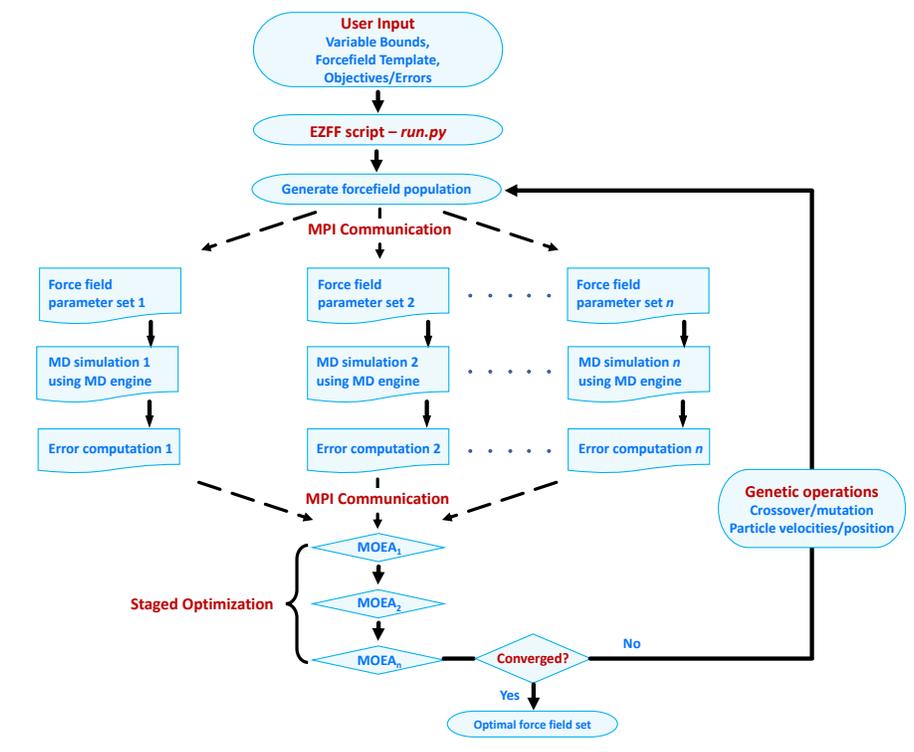

**Figure 1: EZFF forcefield parameterization workflow.** The forcefield parameterization process begins with user specification of forcefield templates and variable bounds in the EZFF script. These are used to generate an ensemble of forcefield candidates, which are evaluated independently by MPI ranks spawned by the master EZFF process. The errors (i.e. objectives) computed by these MPI ranks are communicated back to the master process, which performs genetic operations to spawn the next generation of forcefield candidates.

The inputs are collected together in a single user-defined Python script (run.py in Figure 1), which defines functions for the computation of objectives/errors for the forcefield parameterization, as well as other important properties for global optimization (such as the GA algorithm to be used, population size at each epoch, number of epochs and parallelization scheme to be used). Based on the user-defined values, EZFF generates $n$ valid forcefields based on the template and parameters sampled randomly from the permissible ranges, where $n$ is the population size. These different forcefields are evaluated by the user-chosen MD engines, which are spawned in parallel by mpi4py (or sequentially, if MPI is not available). Each MPI ranks creates its own working directory with necessary input files, executable and further utility files as defined by the user (as defined in the *run.py* script). Individual forcefields are assigned to all available MPI ranks in a round robin fashion. Each MPI rank runs an serial instance of user-defined MD engine such as GULP, LAMMPS, RXMD etc, to produce a simulation corresponding to a specified generated population. The material properties correspond to each forcefield is then compared to their ground truth values to compute the error(s) corresponding to each forcefield in the population. These errors/objectives for each MPI rank is communicated to the main EZFF thread, which uses Platypus and user-defined genetic algorithm to perform crossover and mutation operations and particle displacements to generate the next generation of forcefield candidates for evaluation.

EZFF provides several modules for each stage of this parameterization process (Figure 2) including function definitions to support different tasks required for fitting force fields and implementing the parallel workflow interface with different simulation engines. EZFF is the main module that defines the OptProblem and Algorithm classes for flexible definition of optimization parameters (like error function, genetic algorithms, stopping criterion etc.). Module FFIO defines methods to handle I/O operations on forcefield templates, parameter ranges and EZFF-generated forcefields. LAMMPS and RXMD software for performing MD to evaluate the generated forcefields. These modules include functions to spawn and run these MD engines as well as methods to read data from the execution of these MD programs. In addition, EZFF also includes interfaces to popular Density Functional Theory (DFT) based programs like VASP and QChem to read ground truth values.

The errors module defines several common objectives used to evaluating the quality of forcefields such as atomic positions (including bond lengths and angles), atomic charges, crystal structure and lattice constants, vibration energies, elastic constants, phonon frequencies, bond stretching and dissociation energies.
The utilities module contains functions of generic utility including unit conversion and complex force field template generation such as ReaxFF.

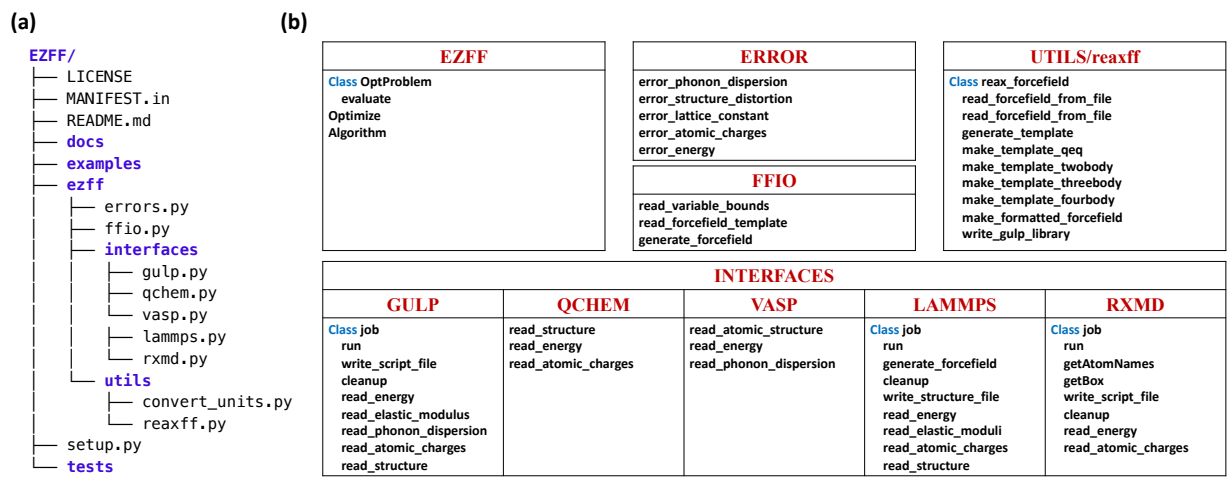

**Figure 2: Organization of EZFF forcefield fitting code.** **(a)** Tree structure of the directory and the files necessary to run EZFF. **(b)** Module diagram showing definitions of various classes and internal organization of flow of control

## 2.1. Installation and compilation
The library is available on the Python Package Index (PyPI) and can be installed using the command

```
pip install EZFF
```
Alternatively, the latest developmental version can be downloaded from the publicly available Github repo at https://github.com/arvk/EZFF and can be compiled by executing setup.py in the root directory of the EZFF tree (Figure 2). The user is responsible for ensuring the installation of the two EZFF dependencies (mpi4py [27, 28] and xtal [29]) separately.

## 2.2. Software functionalities

EZFF provides a lightweight and extensible Python interface that enables, for the first time, serial and parallel multi-objective global parameterization of multiple types of simple and complex reactive and non-reactive forcefields such as ReaxFF, COMB, Stillinger-Weber, Lennard Jones via multiple MD engines. Uniquely, EZFF also allows for the parameterization of hybrid forcefields composed of multiple interatomic interactions with different functional forms.

The modular design of EZFF provides a quick and facile method to change optimization algorithms during forcefield parameterization. This enables strategies such as staged optimization, where diversity preserving genetic algorithms like NSGA-III can be initially employed to more completely sample the parameters space followed by a second stage where other multi-objective optimization schemes like differential evolution can be used to more efficiently converge to local minima in the objective phase space.

Optimization algorithms used in EZFF evolve and keep track of the entire Pareto front at every epoch during optimization. Therefore, we can perform Pareto-frontal uncertainty quantification for forcefields generated by EZFF. This Pareto-frontal breakdown of different forcefields for each epoch provides a natural way to establish one of the primary sources of uncertainty in molecular dynamics simulations – namely the uncertainty in forcefield parameters. This Pareto-frontal uncertainty quantification approach offers an alternative method to estimate the errors in forcefield parameters [11, 30-33], to complement the predominantly Bayesian approaches employed in prior studies [34].

## ILLUSTRATIVE EXAMPLES

We present 2 examples to demonstrate the unique capabilities of EZFF in parameterizing reactive and non-reactive forcefields. In the first example, we demonstrate the parameterization of a hybrid forcefield, consisting of multiple functional forms. The second example covers the optimization of reactive forcefields for modelling accurate metal-polymer interfaces

### Example 1: Optimization of Hybrid Forcefields for Layered Two-dimensional Materials

Two-dimensional and layered materials are being actively investigated for their unique electronic structure and mechanical and transport properties arising out of their quantum confinement along one dimension. In the case of layered transition metal dichalcogenides like $MoS_2$ interatomic interactions can broadly be divided into strong covalent interactions between nearest-neighbor Mo and S atoms, and longer-range van der Waals interactions between $MoS_2$ sheets along the c axis (Figure).

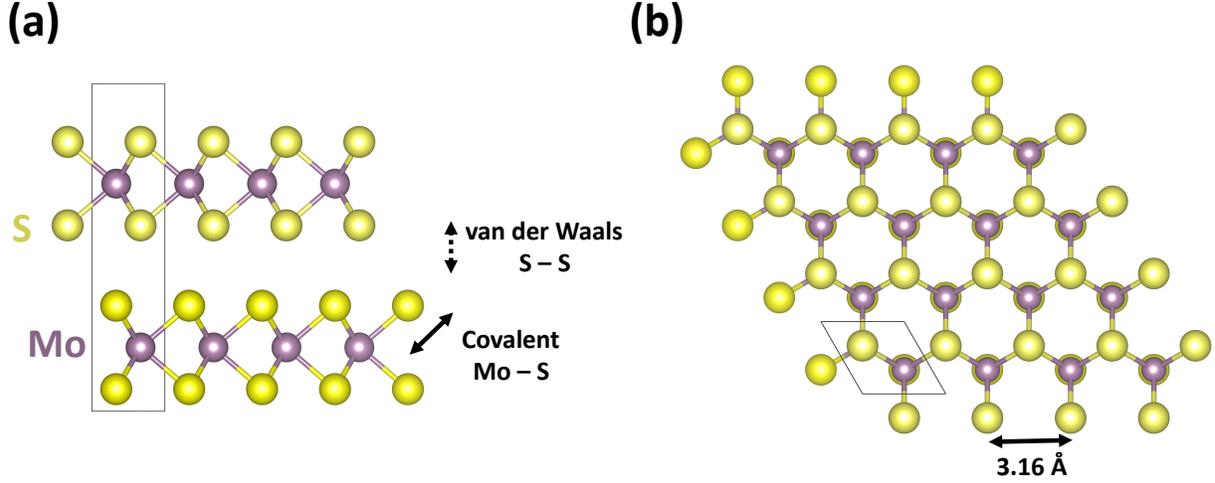

**Figure 3: Crystal structure and interactions in MoS$_2$. (a)** The layered MoS$_2$ crystal is held together by strong in-plane covalent Mo-S interactions and relatively weaker S-S van der Waals interactions. **(b)** The honeycomb crystal structure of MoS2 has a lattice constant of 3.16 Å. The unit cell for MoS$_2$ is indicated by thin black lines.

These interactions are well described by an in-plane Stillinger-Weber interaction between Mo and S atoms combined with out-of-plane Lennard Jones interactions between S atoms in adjacent MoS$_2$ layers. Specifically, this system can be described by a hybrid forcefield that includes:

The total potential energy of the given system of N atoms located at [**r**$_1$, **r**$_2$, …, **r**$_N$] in the SWFF can be written as

$$E_{SW}(\mathbf{r}_1, \mathbf{r}_2, \dots \mathbf{r}_N) = \sum_{i<j} V_2(r_{ij}) + \sum_{i<j<k} V_3(r_{ij}, r_{jk}, \theta_{ijk})$$

where $r_{ij} = |\mathbf{r}_j - \mathbf{r}_i|$. The 2-body term, $V_2$, is defined as

$$V_2(r_{ij}) = A\left(\frac{B}{r_{ij}^4} - 1\right) \exp\left(\frac{\gamma}{r_{ij} - r_{cut}}\right)$$

The two-body term is defined by 3 optimizable parameters, $A$, $B$ and $\gamma$.

The 3-body term, $V_3$ around a central atom $i$ is given by three optimizable parameters, $\lambda$, $\gamma_1$ and $\gamma_2$ and has the following functional form. Geometric parameters, including interaction cut-off distances, $r_{cut}$, $r_{cut1}$ and $r_{cut2}$ and equilibrium angles, $\theta_0$ are held fixed during parameterization.

$$V_3(r_{ij}, r_{ik}, \theta_{ijk}) = \lambda \exp\left(\frac{\gamma_1}{r_{ij} - r_{cut1}} - \frac{\gamma_2}{r_{ik} - r_{cut2}}\right)(\cos\theta - \cos\theta_0)^2$$

Interactions between adjacent MoS$_2$ layers, α and β, are described by Lennard jones interactions between sulfur atoms at r1, r2, r3

$$E_{LJ}\left(\mathbf{r}_1^\alpha, \mathbf{r}_2^\alpha, \ldots \mathbf{r}_1^\beta, \mathbf{r}_2^\beta, \ldots\right) = \sum_{i,j} \varepsilon \left[\left(\frac{\sigma}{\left|r_i^\alpha - r_j^\beta\right|}\right)^{12} - 2 \times \left(\frac{\sigma}{\left|r_i^\alpha - r_j^\beta\right|}\right)^6\right] \forall \ \left|r_i^\alpha - r_j^\beta\right| < r_{cut}^{LJ}$$

This longer-range non-bonded interaction is defined by three optimizable parameters, $\varepsilon$, $\sigma$ and $r_{cut}^{LJ}$.

These parameterizable variables and permissible ranges are defined in the forcefield template and parameter ranges files as shown below.

**Forcefield Template**

```
# E = A.exp(rho/(r-rmax)).(B/r**4 - 1)
# atom1 atom2 A gamma B rmin rmax <3*flags>
sw2
S1    S1   <<A_SS>>    <<G_SS>>    <<B_SS>>    0.00 4.317  0 0 0
Mo    S1   <<A_MoS>>   <<G_MoS>>   <<B_MoS>>   0.00 3.191  0 0 0
Mo    Mo   <<A_MoMo>>  <<G_MoMo>>  <<B_MoMo>>  0.00 4.317  0 0 0
S2    S2   <<A_SS>>    <<G_SS>>    <<B_SS>>    0.00 4.317  0 0 0
Mo    S2   <<A_MoS>>   <<G_MoS>>   <<B_MoS>>   0.00 3.191  0 0 0

# E(three) = lambda * exp(gamma0/(r12-rmax12) + gamma1/(r13-rmax13))(cos-ct0)**2
# atom1 atom2 atom3 lambda theta0 gamma0 gamma1 <rmin12> rmax12 <rmin13> rmax13 <rmin23> rmax23 <4*flags>
sw3
Mo S1 S1 <<L_MoSS>> <<theta_0>> <<G_MoSS>> <<G_MoSS>> 3.191 3.191 4.317 0 0 0 0
S1 Mo Mo <<L_SMoMo>> <<theta_0>> <<G_SMoMo>> <<G_SMoMo>> 3.191 3.191 4.317 0 0 0 0
Mo S2 S2 <<L_MoSS>> <<theta_0>> <<G_MoSS>> <<G_MoSS>> 3.191 3.191 4.317 0 0 0 0
S2 Mo Mo <<L_SMoMo>> <<theta_0>> <<G_SMoMo>> <<G_SMoMo>> 3.191 3.191 4.317 0 0 0 0

# E = A/(r**m) - B/(r**n)
# atom1 atom2 (A B/epsilon sigma) <rmin> rmax <2*flags>
lennard epsilon
S1 S2 <<A_LJ>> <<B_LJ>> <<S_LJ>> 1 0
```

**Parameter Ranges**

```
A_SS    0.9495   1.5825
G_SS    0.34125  0.56875
B_SS   29.025   48.375
A_MoS   5.21475  8.69125
G_MoS   0.273    0.455
B_MoS   6.39    10.65
A_MoMo  3.594    5.99
G_MoMo  0.42375  0.70625
B_MoMo 13.9875  23.3125
L_MoSS  9.13935 15.2323
theta_0 61.5    102.5
G_MoSS  0.828    1.38
L_SMoMo 21.716  36.1933
G_SMoMo 1.78275  2.97125
A_LJ    0.0237   0.0395
B_LJ    2.685    4.475
S_LJ    4.5      7.5
```

**Figure 4: Inputs for EZFF forcefield parameterization.** The forcefield template file is characterized by the presence of named variables (green, enclosed in dual angular brackets, <<>>), which will be replaced by numerical values during the optimization process. The minimum and maximum permissible values for these variables are provided in a separate parameter ranges file, as shown above.

```
1:  import ezff
2:  from ezff.interfaces import vasp, gulp
3:  import numpy as np
4:
5:  bounds = ezff.read_variable_bounds('variable_bounds', verbose=False)
6:  template = ezff.read_forcefield_template('template')
7:
8:  # DEFINE GROUND TRUTHS
9:  gt_disp_GM = vasp.read_phonon_dispersion('/staging/pv/kris658/EZFF/data/GM/band.dat')
10: gt_relax_structure = vasp.read_atomic_structure('/staging/pv/kris658/EZFF/data/POSCAR')
11: gt_c11 = 260.0 #GPa for C11 of MoSe2
12:
13: def my_error_function(variable_values):
14:     myrank = pool.rank
15:     # FOR THE RELAXED STRUCTURE - GM Phonon Dispersion
16:     path = str(myrank)+'/relaxed_GM'
17:     relaxed_job = gulp.job(path=path)
18:     relaxed_job.structure = gt_relax_structure
19:     relaxed_job.forcefield = ezff.generate_forcefield(template, variable_values, FFtype = 'SW')
20:     relaxed_job.options['pbc'] = True
21:     relaxed_job.options['relax_atoms'] = True
22:     relaxed_job.options['relax_cell'] = True
23:     relaxed_job.options['phonon_dispersion'] = True
24:     relaxed_job.options['phonon_dispersion_from'] = '0 0 0'
25:     relaxed_job.options['phonon_dispersion_to'] = '0.5 0.0 0'
26:     relaxed_job.run()  # Submit job and read output
27:     # Read output from completed GULP job and cleanup job files
28:     disp_GM = relaxed_job.read_phonon_dispersion()
29:     md_relaxed_moduli = relaxed_job.read_elastic_moduli()
30:     md_relaxed_structure = relaxed_job.read_structure()
31:     relaxed_job.cleanup()   # FINISH RELAXED JOB
32:     # Compute 4 errors from the GULP job
33:     error_abc, error_ang = ezff.error_lattice_constant(MD=md_relaxed_structure, GT=gt_relax_structure)
34:     a_lattice_error = np.linalg.norm(error_abc[0])    # Error in 'a' lattice constant
35:     c_lattice_error = np.linalg.norm(error_abc[2])    # Error in 'c' lattice constant
36:     md_c11 = md_relaxed_moduli[0][0]
37:     modulus_error_c11 = np.linalg.norm(md_c11 - gt_c11)
38:     phon_error_GM = ezff.error_phonon_dispersion(MD=disp_GM, GT=gt_disp_GM, weights='acoustic')
39:     return [a_lattice_error, c_lattice_error, modulus_error_c11, phon_error_GM]
40:
41: pool = ezff.Pool()
42: problem = ezff.OptProblem(num_errors=4, variable_bounds=bounds, error_function=my_error_function, template=template)
43: algorithm = ezff.Algorithm(problem, 'NSGAIII', population = 256, pool = pool)
44: ezff.optimize(problem, algorithm, iterations = 100)
45: pool.close()
```

**Figure 5:** Complete script (run.py) for preforming parameterization of a hybrid Stillinger-Weber and Lennard-Jones forcefield using EZFF

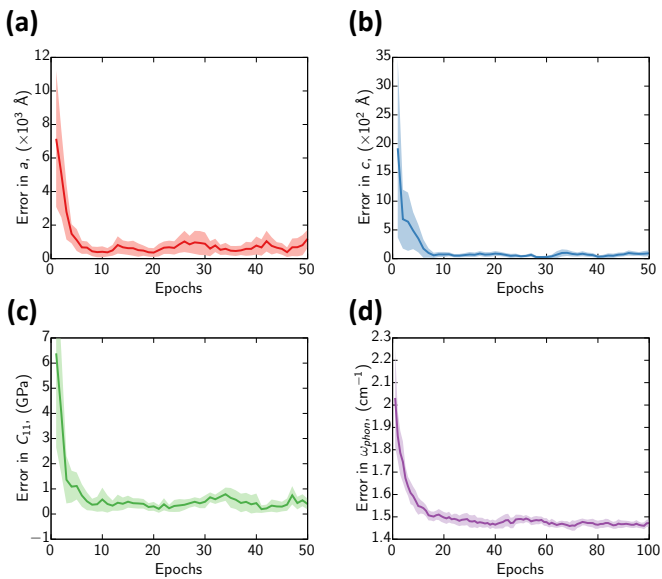

**Figure 6: Quality of forcefields during optimization.** (a-d) show the computed error for each of the four objectives used for parameterizing the hybrid Stillinger-Weber and Lennard-Jones forcefield. At each epoch, the mean (dark line) and standard deviation (light fill) of the 20 best forcefields on the Pareto front are plotted. The NSGA2 algorithm converges quickly to produce good forcefields within 50 epochs.

Figure 6 shows that by using NSGA-III along with EZFF quickly identifies a set of forcefields that simultaneous optimizes all 4 objectives considered in this example.

### Example 2: Optimization of ReaxFF forcefields for Al-polymeric materials using RXMD

Understanding the electronic properties at metal-organic interfaces are becoming increasingly crucial as electronic devices like batteries and capacitors move towards smaller scales and higher efficiency. But the experimental exploration of interfacial electronic properties is mired with challenges owing to their immense chemical and morphological complexity [35-37]. To side-step this difficulty, the community currently uses first-principles computations on highly simplified models of interfacial structures to access these properties [38, 39]. But, the understanding of electronic processes is currently limited by the ideal nature of the interfacial structures used in these simulations. While generation of realistic interface structures using first-principle based methods remains intractable due its complexity, structure generation using MD simulations utilizing reactive forcefields like ReaxFF [4, 40] remains a feasible option. Here, we attempt to generate a ReaxFF forcefield using EZFF which can accurately capture the interaction between Aluminum (which is a common electrode) and C-H-O based organic molecules/polymers to facilitate easy creation of realistic Al-organic interfaces.

ReaxFF is a class of semi-empirical bond-order-based forcefields for describing reactive dynamics involving bond breaking and formation and are well suited to describe highly heterogeneous material systems. ReaxFF forcefields are composed of several hundred parameterizable variables that describe various 2-body, 3-body and 4-body interactions between different atomic species, which makes global optimization of these potentials highly challenging. Recently, Hong and van Duin parameterized a new ReaxFF forcefield for Al/C/H/O materials against interaction energies between organic radicals and Al [41]. However, this work fails to capture the interaction energy between fully saturated organic molecules and the Al surface correctly resulting in unrealistic interface structures. Therefore, we use EZFF to reparameterize the ReaxFF forcefield from Hong and van Duin to better reproduce interaction energies between an Al (111) surface and two representative saturated organic molecules, $CH_4$ and $C_2H_6$. Specifically, EZFF is used to parameterize only variables controlling the Al-C and Al-H 2-body interactions to most accurately reproduce the DFT-computed interaction energy between Al and $CH_4$/$C_2H_6$.

Figure 7 shows the full run.py script to perform forcefield parameterization using the RXMD [17] as the MD engine to evaluate the quality of different ReaxFF forcefield. The script uses the Indicator-Based Evolutionary Algorithm (IBEA) to optimize the ReaxFF parameters over 100 epochs against two objectives – deviations from DFT-computed energies for the Al-$CH_4$ and Al-$C_2H_6$ systems respectively.

```python
1:  import ezff
2:  from ezff.interfaces import gulp, vasp
3:  from ezff.utils.reaxff import reax_forcefield
4:
5:  # Define ground truths
6:  structure_ch4 = vasp.read_atomic_structure('ground_truths/ch4')
7:  energy_ch4 = vasp.read_energy('ground_truths/ch4')
8:  structure_c2h6 = vasp.read_atomic_structure('ground_truths/c2h6')
9:  energy_c2h6 = vasp.read_energy('ground_truths/c2h6')
10:
11: def my_error_function(rr):
12:     path = str(pool.rank)
13:
14:     # Calculate CH4 structure
15:     ch4_job = gulp.job(path = path)
16:     ch4_job.structure = structure_ch4
17:     ch4_job.forcefield = ezff.generate_forcefield(template, rr, FFtype = 'reaxff')
18:     ch4_job.options['pbc'] = True
19:     ch4_job.options['relax_atoms'] = False
20:     ch4_job.options['relax_cell'] = False
21:     # Run gulp calculation
22:     ch4_job.run(command='gulp')
23:     # Read output from completed GULP job and clean-up
24:     ch4_md_energy = ch4_job.read_energy()
25:     ch4_job.cleanup()
26:
27:     # Calculate C2H6 structure
28:     c2h6_job = gulp.job(path = path)
29:     c2h6_job.structure = structure_c2h6
30:     c2h6_job.forcefield = ezff.generate_forcefield(template, rr, FFtype = 'reaxff')
31:     print(c2h6_job.forcefield[0])
32:     c2h6_job.options['pbc'] = True
33:     c2h6_job.options['relax_atoms'] = False
34:     c2h6_job.options['relax_cell'] = False
35:     # Run gulp calculation
36:     c2h6_job.run(command='gulp')
37:     # Read output from completed GULP job and clean-up
38:     c2h6_md_energy = c2h6_job.read_energy()
39:     c2h6_job.cleanup()
40:
41:     # Calculate errors
42:     ch4_error = ezff.error_energy(ch4_md_energy, energy_ch4-energy_ch4[-1], weights = 'uniform')
43:     c2h6_error = ezff.error_energy(c2h6_md_energy, energy_c2h6-energy_c2h6[-1], weights = 'uniform')
44:
45:     return [ch4_error, c2h6_error]
46:
47:
48: pool = ezff.Pool()
49:
50: if pool.is_master():
51:     # Generate forcefield template and variable ranges
52:     FF = reax_forcefield('AlCHO.ff')
53:     FF.make_template_twobody('Al','C')
54:     FF.make_template_twobody('Al','H')
55:     FF.generate_templates()
56:
57: # Read template and variable ranges
58: bounds = ezff.read_variable_bounds('param_ranges', verbose=False)
59: template = ezff.read_forcefield_template('ff.template.generated')
60:
61: problem = ezff.OptProblem(num_errors=2, variable_bounds=bounds, error_function=my_error_function, template=template)
62: algorithm = ezff.Algorithm(problem, 'NSGAII', population = 239, pool = pool)
63: ezff.optimize(problem, algorithm, iterations = 100, write_forcefields = 1)
64: pool.close()
```

**Figure 7:** Complete script (run.py) for preforming parameterization of a ReaxFF forcefield for the polymer-Al system using EZFF

Figures 8a and 8b show that IBEA converges rapidly within 100 epochs producing optimal forcefields that replicate the adsorption energy profiles for both $CH_4$ and $C_2H_6$ molecules and Figure 9 shows that the adsorption energy profiles from the optimized forcefields are in much better agreement with the DFT values than those generated from the original forcefield from Hong and van Duin.

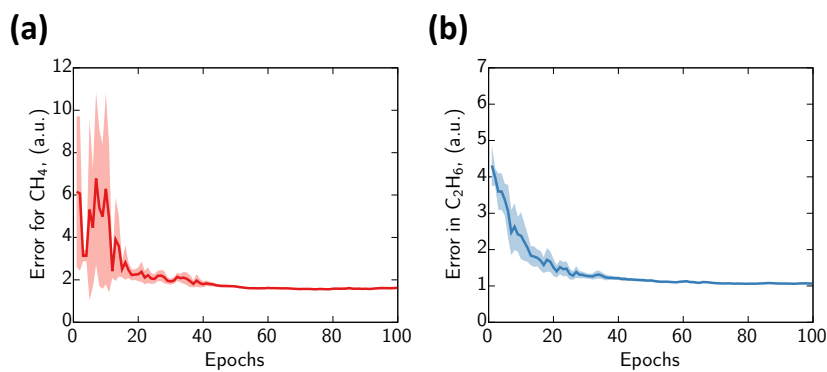

**Figure 8: Quality of ReaxFF forcefields during genetic optimization.** (a) and (b) show the computed error in the adsorption energy profile respectively for the Al-CH$_4$ and Al-C$_2$H$_6$ systems. At each epoch, the mean (dark line) and standard deviation (light fill) of the 20 best forcefields on the Pareto front are plotted. The IBEA algorithm employed here converges quickly to produce good forcefields within 50 epochs.

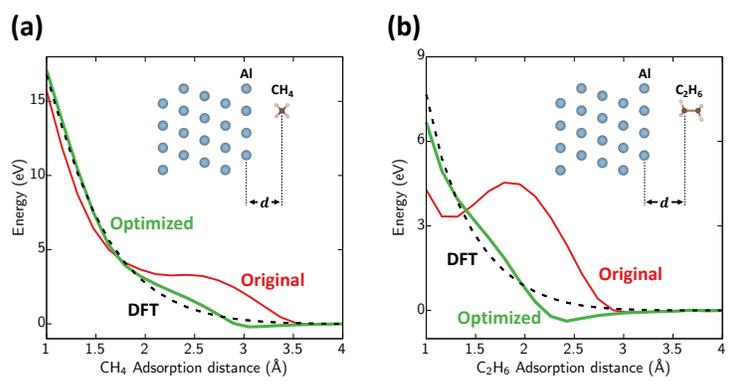

**Figure 9: Adsorption energy profile of optimized ReaxFF forcefield.** Al-molecule interaction energies from the optimized forcefield (100$^{th}$ epoch) are in closer agreement with the DFT ground truth values than those from the original forcefield from Hong and van Duin.

## 3. IMPACT

Parameterization of interatomic forcefields is a highly time-consuming and cumbersome process, whose complexity, along with the lack of general best-practices guidelines has led the process of forcefield construction to be considered an 'art'. EZFF attempts to improve the process of forcefield parameterization by providing a simple workflow, in an easy-to-understand scripting language to optimize a wide range of empirical forcefields of varying levels of complexity. The highly parallelized parameterization process enables rapid prototyping and testing of multiple forcefields before performing production molecular dynamics simulations. The parameterization of hybrid forcefields opens doors for the direct parameterization of interatomic interactions for highly heterogeneous material systems, including those containing interfaces between two distinct phases. This would foster an ensemble of exploratory studies into the rich and largely unexplored space of interfacial properties which are rather exotic compared to bulk materials.

## 4. CONCLUSIONS

In this paper, we described EZFF, a lightweight and flexible Python library for multi-objective global parameterization of different types of interatomic forcefields for molecular dynamics simulations. The highly parallelized and scalable framework will enable quick prototyping of several forcefield function forms, as well as hybrid forcefields composed of multiple interatomic interactions. EZFF also admits staged optimization strategies using multiple optimization algorithms for generating high-quality forcefields with built-in Pareto-frontal uncertainty quantification, thus greatly simplifying the currently cumbersome process for construction and validation of forcefields.

The EZFF codebase is meant to continuously evolve in future releases, welcoming suggestions and contributions from users. Future versions will include capability to use dynamic properties (mean square displacements, and various correlation functions) as objectives in the parameterization process, enabling fitting of forcefields to dynamic material properties.

## 5. ACKNOWLEDGMENTS

*This work was supported by the Office of Naval Research through a Multi-University Research Initiative (MURI) grant (N00014-17-1-2656). This work used the Extreme Science and Engineering Discovery Environment (XSEDE), which is supported by National Science Foundation grant number ACI-1548562 and The Partnership for Advance Computing Environment (PACE) at Georgia Institute of Technology.*

*Table 1 – Code metadata (mandatory)*

| Nr | Code metadata description | |
|----|---------------------------|---|
| C1 | Current code version | *v0.9.4* |
| C2 | Permanent link to code/repository used of this code version | *https://github.com/arvk/EZFF/* |
| C3 | Legal Code License | *MIT License* |
| C4 | Code versioning system used | *git* |
| C5 | Software code languages, tools, and services used | *Python* |
| C6 | Compilation requirements, operating environments & dependencies | *Message Passing Interface (MPI) libraries* |
| C7 | If available Link to developer documentation/manual | *https://ezff.readthedocs.io/en/latest/* |
| C8 | Support email for questions | *cacs@usc.edu* |

*Table 2 – Software metadata (optional)*

| Nr | (Executable) software metadata description | |
|----|--------------------------------------------|---|
| S1 | Current software version | *0.9.4* |
| S2 | Permanent link to executables of this version | *https://github.com/arvk/EZFF/archive/v0.9.4.zip* |
| S3 | Legal Software License | *MIT License* |
| S4 | Computing platforms/Operating Systems | *Linux, OSX* |
| S5 | Installation requirements & dependencies | *Message Passing Interface (MPI) libraries* |
| S6 | If available, link to user manual - if formally published include a reference to the publication in the reference list | *https://ezff.readthedocs.io/en/latest/* |
| S7 | Support email for questions | *cacs@usc.edu* |